\documentclass[twoside,twocolumn,9pt]{article}
\usepackage{extsizes}
\usepackage[super,sort&compress,comma]{natbib} 
\usepackage[version=3]{mhchem}
\usepackage[left=1.5cm, right=1.5cm, top=1.785cm, bottom=2.0cm]{geometry}
\usepackage{balance}
\usepackage{mathptmx}
\usepackage{sectsty}
\usepackage{graphicx} 
\usepackage{lastpage}
\usepackage[format=plain,justification=justified,singlelinecheck=false,font={stretch=1.125,small,sf},labelfont=bf,labelsep=space]{caption}
\usepackage{float}
\usepackage{fancyhdr}
\usepackage{fnpos}
\usepackage[english]{babel}
\addto{\captionsenglish}{%
  
}
\usepackage{array}
\usepackage{droidsans}
\usepackage{charter}
\usepackage[T1]{fontenc}
\usepackage[usenames,dvipsnames]{xcolor}
\usepackage{setspace}
\usepackage[compact]{titlesec}
\usepackage{hyperref}
\usepackage{wrapfig}
\usepackage{textcomp}
\usepackage{eurosym}

\usepackage{epstopdf}%This line makes .eps figures into .pdf - please comment out if not required.

\definecolor{cream}{RGB}{222,217,201}

\begin{document}

\pagestyle{fancy}
\thispagestyle{plain}
\fancypagestyle{plain}{
%%%HEADER%%%
\renewcommand{\headrulewidth}{0pt}
}
%%%END OF HEADER%%%

%%%PAGE SETUP - Please do not change any commands within this section%%%
\makeFNbottom
\makeatletter
\renewcommand\LARGE{\@setfontsize\LARGE{15pt}{17}}
\renewcommand\Large{\@setfontsize\Large{12pt}{14}}
\renewcommand\large{\@setfontsize\large{10pt}{12}}
\renewcommand\footnotesize{\@setfontsize\footnotesize{7pt}{10}}
\renewcommand\scriptsize{\@setfontsize\scriptsize{7pt}{7}}
\makeatother

\renewcommand{\thefootnote}{\fnsymbol{footnote}}
\renewcommand\footnoterule{\vspace*{1pt}% 
\color{cream}\hrule width 3.5in height 0.4pt \color{black} \vspace*{5pt}} 
\setcounter{secnumdepth}{5}

\makeatletter 
\renewcommand\@biblabel[1]{#1}            
\renewcommand\@makefntext[1]% 
{\noindent\makebox[0pt][r]{\@thefnmark\,}#1}
\makeatother 
\renewcommand{\figurename}{\small{Fig.}~}
\sectionfont{\sffamily\Large}
\subsectionfont{\normalsize}
\subsubsectionfont{\bf}
\setstretch{1.125} %In particular, please do not alter this line.
\setlength{\skip\footins}{0.8cm}
\setlength{\footnotesep}{0.25cm}
\setlength{\jot}{10pt}
\titlespacing*{\section}{0pt}{4pt}{4pt}
\titlespacing*{\subsection}{0pt}{15pt}{1pt}
%%%END OF PAGE SETUP%%%

%%%FOOTER%%%
\fancyfoot{}
%\fancyfoot[LO,RE]{\vspace{-7.1pt}\includegraphics[height=9pt]{head_foot/LF}}
%\fancyfoot[CO]{\vspace{-7.1pt}\hspace{13.2cm}\includegraphics{head_foot/RF}}
%\fancyfoot[CE]{\vspace{-7.2pt}\hspace{-14.2cm}\includegraphics{head_foot/RF}}
%\fancyfoot[RO]{\footnotesize{\sffamily{1--\pageref{LastPage} ~\textbar  \hspace{2pt}\thepage}}}
\fancyfoot[LE,RO]{\footnotesize{{\thepage~\textbar}}}
\fancyhead{}
\renewcommand{\headrulewidth}{0pt} 
\renewcommand{\footrulewidth}{0pt}
%\setlength{\arrayrulewidth}{1pt}
%\setlength{\columnsep}{6.5mm}
%\setlength\bibsep{1pt}
%%%END OF FOOTER%%%

%%%FIGURE SETUP - please do not change any commands within this section%%%
%\makeatletter 
%\newlength{\figrulesep} 
%\setlength{\figrulesep}{0.5\textfloatsep} 

%\newcommand{\topfigrule}{\vspace*{-1pt}% 
%\noindent{\color{cream}\rule[-\figrulesep]{\columnwidth}{1.5pt}} }

%\newcommand{\botfigrule}{\vspace*{-2pt}% 
%\noindent{\color{cream}\rule[\figrulesep]{\columnwidth}{1.5pt}} }

%\newcommand{\dblfigrule}{\vspace*{-1pt}% 
%\noindent{\color{cream}\rule[-\figrulesep]{\textwidth}{1.5pt}} }

%\newcommand{\un}[1]{\, \mathrm{#1}}

\definecolor{brightmaroon}{rgb}{0.76, 0.13, 0.28}

%\makeatother
%%%END OF FIGURE SETUP%%%

%%%TITLE AND AUTHORS%%%
%\twocolumn[
%  \begin{@twocolumnfalse}
%{\includegraphics[height=30pt]{head_foot/journal_name}\hfill\raisebox{0pt}[0pt][0pt]{\includegraphics[height=55pt]{head_foot/RSC_LOGO_CMYK}}\\[1ex]
%\includegraphics[width=18.5cm]{head_foot/header_bar}}\par
%\vspace{1em}
%\sffamily
%\begin{tabular}{m{4.5cm} p{13.5cm} }

%\includegraphics{head_foot/DOI} & \noindent\LARGE{\textbf{Beyond power density: unexpected scaling laws in scale up of characterization of reverse-electro-dialysis membranes.}} \\%Article title goes here instead of the text "This is the title"
% & \vspace{0.3cm} \\

% & \noindent\large{Timothée Derkenne $^{a}$, Annie Colin$^{a}$ and Corentin Tregouet$^{*,a}$}
% \\%Author names go here instead of "Full name", etc.

%\includegraphics{head_foot/dates} & \\

%\end{tabular}

% \end{@twocolumnfalse} \vspace{0.6cm}

%  ]

\twocolumn[
\begin{@twocolumnfalse}

\noindent\LARGE{\textbf{Beyond power density: unexpected scaling laws in scale up of characterization of reverse-electro-dialysis membranes.}}\\

\noindent\large{Timothée Derkenne $^{a}$, Annie Colin$^{a}$ and Corentin Tregouet$^{*,a}$}\vspace{0.6cm}  

%%%ABSTRACT%%%%
\section*{Abstract}

Blue energy represents a large reservoir of renewable osmotic energy that can be converted into electricity by reverse electrodialysis (RED). This method is based on ion-exchange membrane. Before large scale production, these membranes are compared on very small samples on the basis of the power they enable to produce per unit area. Through a systematic study of the effect of the membrane size on the power density, we show experimentally that for classical measurement cells, the power density strongly varies with the size of the membrane: the smaller the membrane, the higher the power density. The results are explained by a theoretical modeling which describes the effect of the access resistance at the scale of the membrane. Based on this work, a few recommendations are formulated to perform scalable and meaningful measurements of membrane resistance and power density. 
    
\vspace{0.6cm}  

%\sffamily{\textbf{Abstract here.}}

\textbf{\underline{Key words}: blue energy, nanofluidics, reverse electrodialysis, access resistance}

\end{@twocolumnfalse} \vspace{1.6cm}
]
%\includegraphics[width=7cm]{Graphical_abstract.pdf}

%%%END OF TITLE AND AUTHORS%%%

%\noindent\LARGE{\textbf{Beyond power density: unexpected scaling laws in scale up of characterization of reverse-electro-dialysis membranes.}}

%%%FONT SETUP - please do not change any commands within this section
%\renewcommand*\rmdefault{bch}\normalfont\upshape
\rmfamily
\section*{}
\vspace{-1cm}

%%%FOOTNOTES%%%

\footnotetext{\textit{$^{a}$~ESPCI Paris, PSL Research University, MIE-CBI, CNRS UMR 8231, 10, Rue Vauquelin, F-75231 Paris Cedex 05, France.}}

\footnotetext{\textit{$^{*}$~Corresponding author: corentin.tregouet@espci.psl.eu}}

%Please use \dag to cite the ESI in the main text of the article.
%If you article does not have ESI please remove the the \dag symbol from the title and the footnotetext below.
%\footnotetext{\dag~Electronic Supplementary Information (ESI) available: [details of any supplementary information available should be included here]. See DOI: 00.0000/00000000.}
%additional addresses can be cited as above using the lower-case letters, c, d, e... If all authors are from the same address, no letter is required

%\footnotetext{\ddag~Additional footnotes to the title and authors can be included \textit{e.g.}\ `Present address:' or `These authors contributed equally to this work' as above using the symbols: \ddag, \textsection, and \P. Please place the appropriate symbol next to the author's name and include a \texttt{\textbackslash footnotetext} entry in the the correct place in the list.}

%%%END OF FOOTNOTES%%%

%%%ABSTRACT%%%%
%\section*{abstract}

%\sffamily{\textbf{Abstract here.}}

%\textbf{\underline{Key words}:}

%citation type for the abstract [Surname \textit{et al., Journal Title}, 2000, \textbf{35}, 3523]
%\end{abstract}
%%%END OF ABSTRACT%%%%

%%% Graphical ABSTRACT%%%%

%\rmfamily %Please do not remove this line.

%%%MAIN TEXT%%%%

\section{Introduction}

% Blue energy
Global demand on energy is growing due to the urging necessity to switch to decarbonated energy production. Blue energy (BE), by harvesting osmotic energy from salt gradients, typically where fresh-water rivers enter the salty ocean, has the potential to be a valuable new source of energy \cite{Logan2012MembraneBased}. Other sources of salt gradient can arise from industrial or domestic waste water, or dedicated engineered fluids \cite{yipSalinityGradientsSustainable2016, PascualWastedHeat}. It is indeed possible to generate spontaneous ionic flows through nanoporous membranes and collect this electrical energy: this process of energy harvesting is called reverse electro dialysis (RED). %BE’s worldwide available power is estimated to be 1 TW (8800 TWh/year)\cite{Logan2012MembraneBased,Watersalinationsourceofenergy}. 
%\color{orange} 
%le chiffre est un peu exagéré .\color{black} 
Considering the suitability, sustainability and reliability of the exploitation \cite{yipSalinityGradientsSustainable2016,Logan2012MembraneBased,wuFluidicsEnergyHarvesting2023a,elimelechFutureSeawaterDesalination2011,marbachOsmosisMolecularInsights2019b}, the blue energy power that could realistically be harvested can be estimated as 625 TWh/year \cite{Essalhi2023potential,AlvarezSilva2016practicalglobal}, which corresponds to $2.2\%$ share of global electricity consumption in 2022 \cite{Energyinstitute2023}. 

%\color{black}
%\cite{Ramon2011Estimation}
% Need for power vs quality and price of membrane

Most harvesting methods are based on ion-exchange membranes containing typically nanometric pores (1-nm pores). The main origin of cost to harvest this energy is the planar membranes. As a consequence, technologies are benchmarked in terms of power per unit area of membrane. Real-scale prototype plants reach about $1 \mathrm{\,W/m^2}$ of membrane \cite{VEERMAN20097}.The profitability threshold  is estimated at $5 \mathrm{\,W/m^2}$~~\cite{Jia2014Blue} (for a membrane cost around $2\$$/m$^2$). At present, the price of the cheapest selective membranes in SPEEK is still higher than 10 euros per m$^{2}$ \cite{yuan2022low}. The technology is therefore still a long way from economic viability, and needs to be improved.
\color{black}

% Rescaling

%\color{orange} 
%je trouve la formulation et la justification de la necessité du scaling un peu bizarre. 

Over the last ten years or so, numerous studies have focused on the synthesis of new membranes and their improvement to overcome this barrier \cite{Hong2015PotentialIEM,Mei2018RecentDevelopment,Laucirica2021nanofluidic}.
Most of the innovative membranes are characterized in the labs by the power per unit area (power density) on devices with a membrane area of the order of $10 \mathrm{\,\mu m^2}$, assuming that for given fluids, the power density depends only on the membrane material and thickness, and will be poorly affected by the scale-up. This is justified by the assumption that the measured  resistance in the experimental device corresponds to the one of the  most resistive object, i.e. the membrane. The power density of the membrane is thus calculated by dividing the harvested power by its area.

%Beyond the management liquid fluxes, which are classical problem of chemical engineering, 
%The question of scaling up the membrane area and the ion flux through it raises many questions.

%\color{black}

%Scaling up from a unique nanopore to a collection of nanopore or nanoporous membranes already raises numerous questions, since the power is far from being proportionnal to the number of channels \cite{Gao2019Understanding,Yazda2021HighOsmotic}. The collective effects between the neighboring nanopores that prevent the extrapolation form a unique nanopore to nanoporous membranes are not fully understood yet.

%More specifically, questions remain open regarding a size effect on the maximum harvested power density, which.

%However, as for most technological novelties, scale up is challenging. Beyond the management liquid fluxes, which are classical problem of chemical engineering, the question of scaling up the membrane area and the ion flux through it raises many questions. While any prototype has a membrane area of at least $1 \mathrm{\,cm^2}$, most of the innovative membranes are characterized in the labs on devices with a membrane area of the order of $10 \mathrm{\,\mu m^2}$. This is justified by the assumption that the power density  (power per unit area of membrane) is intrinsically linked to the membrane, and is not affected by the scale up. 

%\color{orange}
Recent literature reviews show a correlation between membrane area and power density and call this hypothesis into question\cite{Yang2022CrosslinkedNanocellulose,Lin2023temperaturegated,Gao2023bioinspired,Pan2023TowardScalable,Lin2023Essence}. Some of the phenomena behind these measurements are easy to understand. The ionic resistance of the experimental device is due not only to the membrane, but also to its surroundings, in particular the electrolyte reservoirs. As an example, when the membrane size is increased, the membrane resistance get smaller and can even become smaller than the resistance of the reservoirs (independent of the membrane area). In such a situation, as the size of the membrane would increase, in the situation where the surroundings are unchanged, the power recovered would remain constant and the density power (power by  membrane area unit) would decrease. Other possible effects are more complex. The spatial distribution of ions in the electrolyte can be affected by the selectivity of the membrane, the geometry of the device i.e. membrane size, electrode size, distance between electrodes and membranes. 

%\color{black}
%\color{orange}
%j'enleverai ca
%So far, due to the absence of standard measurement protocols, membranes are characterized of various membrane sizes.
%\color{black}
% diversity of power obtained (insist more small vs large membrane, maybe mention monopore vs multi pore and what area is considered in these 2 cases)

%However, some papers reported that the power density for a given membrane might depend on the membrane area \cite{Yang2022CrosslinkedNanocellulose,Lin2023temperaturegated} Moreover, due to the absence of standard measurement protocols, membranes are characterized of various membrane sizes, and recent literature reviews show a correlation between membrane area and power density \cite{Yang2022CrosslinkedNanocellulose,Pan2023TowardScalable}.

In this paper, the relation between membrane area and power density is systematically investigated for a given commercial membrane (Nafion 115) in a standard electro-chemical cell.
In a first part (section \ref{resistanceSec}), the cell resistance is measured with respect to the membrane area with the same salt concentration on both sides of the membrane. In a second part (section \ref{powerSec}), resistance and produced power are measured with different salt concentrations on both sides of the membrane. These results are then discussed (section \ref{DiscussionSec}) in comparison with the literature, and implications are drawn for further research.

We mainly highlight two phenomena: 1- the conductance is not proportional to the membrane surface, but to its square root for classical electro-chemical cell dimensions; 2- as a consequence, the maximum power density strongly decreases with membrane size. As a result, extrapolation of power density from measurements on small membranes is flawed and overly optimistic.

%Experimental results show that in a standard measurement geometry where electrodes have a fixed size and are larger than the membrane, the power density decreases with the membrane area with a power law \textonehalf . We also demonstrate that this power law is well described by a theoretical model based on a geometrical phenomenon well known for nanopores. The power density is usually considered as independent of the system size and hence unchanged during scale up. Therefore, these results assess the necessity of defining measurement standards in order to measure membrane performances that can be extrapolated. Finally we formulate recommendations to reach this objective.

\section{Results}

\begin{figure}
\centering
    \includegraphics[width=\linewidth]{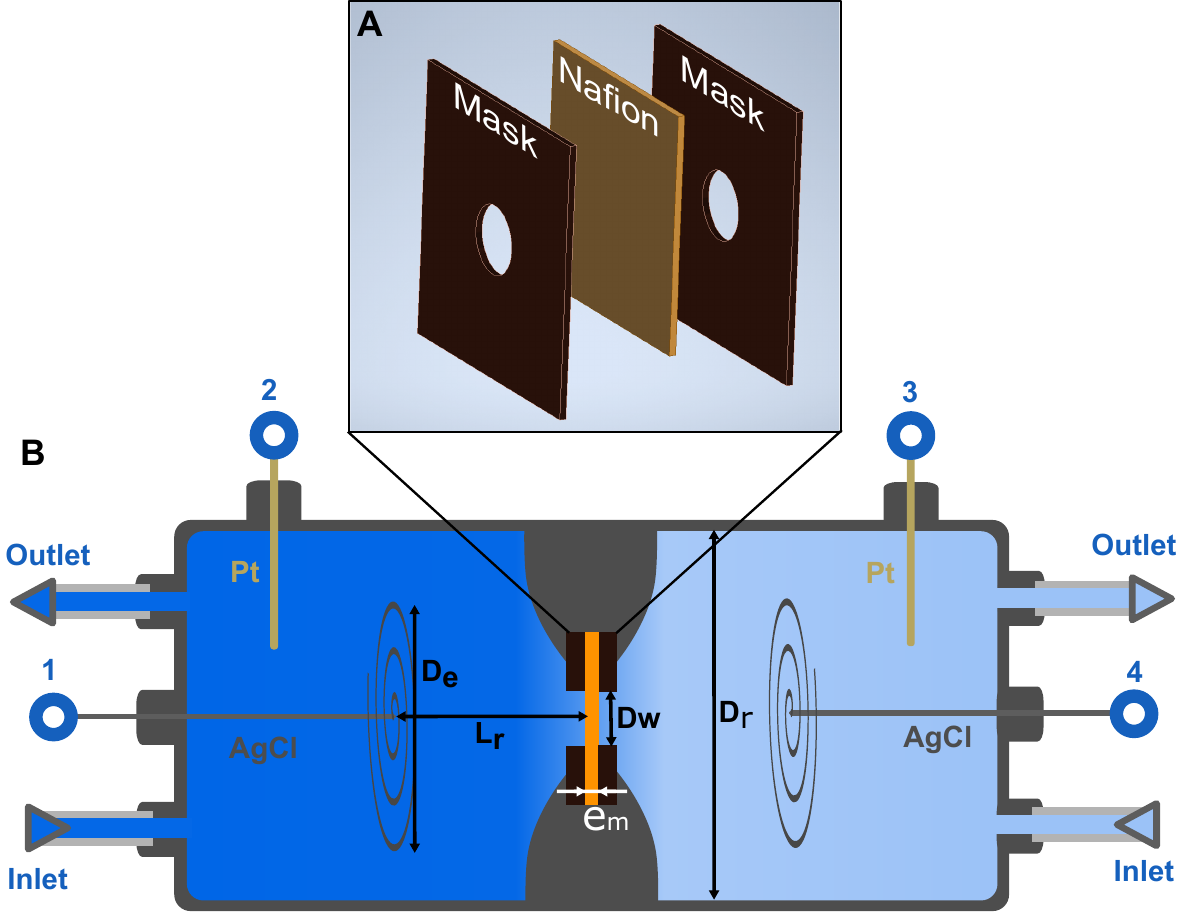}
    \caption{Experimental set up used for all the measurements. (a) 3D view of the membrane (yellow square) mounted between two mylar masking windows (red square with a circular hole). The size of this hole defines the effective $S_{membrane}$ taken into account.(b) Schematic of the cell with the two reservoirs of salt solutions, with inlet and outlet. The two silver chloride electrodes made of a wire spiral (labels 1 and 4). The membrane (orange) fixed in a mounting bracket with a masking window on each side (black). Two platinum wires (labels 2 and 3) are placed behind the silver chloride electrodes. Electric circuit are presented SI. Geometrical parameters are indicated in Table \ref{table_dimension}.
    \label{experimental_setup}}
\end{figure}

An electro-chemical cell (Figure \ref{experimental_setup}) was designed to measure Nafion-membrane characteristics (Nafion 115). The Nafion membrane is clamped between two masking windows to vary the membrane effective surface from $10^{-2}$ to $10^2 \, \mathrm{mm^2}$. This assembly (membrane and masking windows) divides the electro-chemical cell into two reservoirs. As shown in Figure \ref{experimental_setup}, from each side come the water inlet and outlet, the silver-chloride electrodes and the platinum electrodes (more details in the Methods section)

Cell electrical resistance $R_{cell}$, open-circuit voltage $E_{OCV}$ and maximum power $P_{max}$ were measured for varying membrane surface and salt solutions. Salt solutions (labeled 1, 2 and 3) consist of potassium chloride of varying concentrations: solution 1 has a concentration $c_1 = 1 \mathrm{\,g/L}$ ($1.7 \cdot 10^{-2} \mathrm{\, M}$) and a conductivity $\sigma_1= 1850 \mathrm{\, \mu S/cm}$; solution 2 has a concentration close to $c_2 = 10 \mathrm{\,g/L}$ ($1.7 \cdot 10^{-1} \mathrm{\, M}$) such as $\sigma_2 = 10.\sigma_1$; solution 3 has a concentration close to $c_3 = 100 \mathrm{\,g/L}$ ($1.7 \mathrm{\, M}$) such as $\sigma_3 = 100.\sigma_1$ (more details in Methods section). 

%Briefly, a cation-selective membrane (Nafion 115) is placed in an electro-chemical cell to seperate two reservoirs containing salt solutions. Masking windows are placed against the membrane to tune the effective surface area of membrane from $10^{-2}$ to $10^2 \, \mathrm{mm^2}$. 

Electrical measurements are performed using a 2-electrode configuration and a 4-electrode configuration. With two electrodes only, the potential is measured at electrodes in which current is flowing. In some conditions (discussed in the Method section), these electrodes are therefore submitted to polarization at the interface, inducing an excess resistance \cite{Bazant2005currentvoltage} which depends on salt concentration. Consequently, at high concentration ($c_2$ and $c_3$), when the bulk resistance is low, the electrode polarization resistance is of the same order of magnitude as the total resistance. To overcome this issue, the 4-electrode configuration enables a voltage measurement on electrodes through which there is no current, and hence no effect of the polarization resistance. In the conditions where there is no effect of electrode polarisation, the 2-electrode configuration is used. More details on the setup and the measurement protocol are given in the method section.

\subsection{Resistance} 
\label{resistanceSec}
\subsubsection{Experimental results}

In this first section, resistance measurements are performed for symmetrical configurations (same salt concentration in the two reservoirs, and hence no gradient).

% gross results
% discussion of concentration effect
% slope -1/2

\begin{figure}[h]
\centering
    \includegraphics[width=1\linewidth]{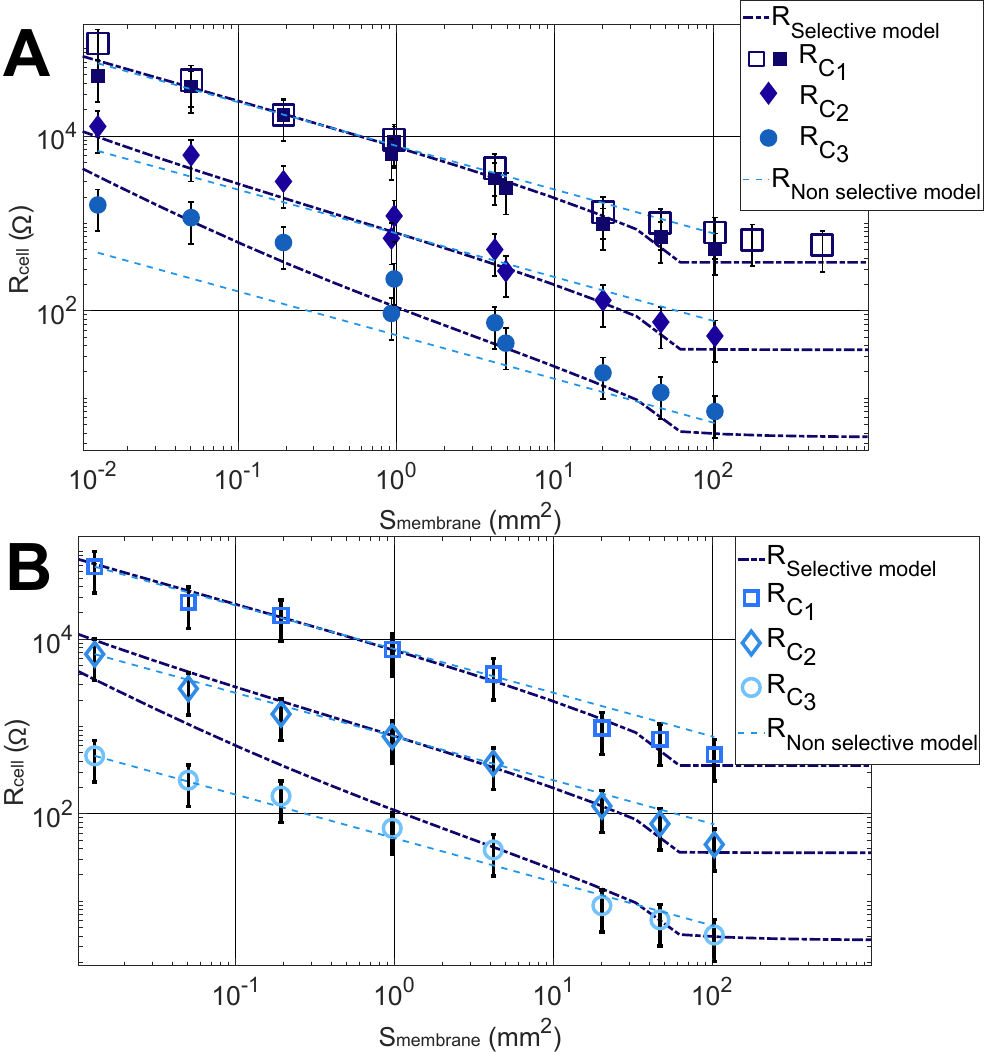}
    \caption{Measured electrical resistances (${R_{cell}}$) with respect to the membrane area ($S_{membrane}$) for varying geometries in symmetrical configurations. (a) With a Nafion membrane (solid symbols): the measured resistance follows the model developed by Green \textit{et al} \cite{Green2014EffectGeometry,Green2016Interplay} (black dashed lines). Open symbols are for a different geometry ($L_2$ instead of $L_1$), as expected no influence of the reservoir length is observed. (b) Without membrane (open symbols): resistance follows the model which shows a trend in $1/\sqrt{S_{membrane}}$ (light blue dashed line).
    Dashed lines from the model are the same in A, B and the same as in Figure \ref{resistance_th}. 
    \label{resistance}}
\end{figure}

In a first step, the cell resistance is measured with solution 1 ($c_1 = 1 \mathrm{\,g/L}$ ($1.7 \cdot 10^{-2} \mathrm{\, M}$) ) for various membrane areas. As shown in Figure \ref{resistance}, these measurements yield a decreasing resistance with respect to the membrane area, which can be approximated by a power law $S_{membrane}^{-1/2}$. This scaling is a very good agreement with the measurements of Lin et al. \cite{Lin2023Essence}.

The same scaling law is obtained with solution 2 ($c_2 = 10 \mathrm{\,g/L}$ ($1.7 \cdot 10^{-1} \mathrm{\, M}$) ), with a resistance 10 times lower. For larger concentration ($c_3= 100 \mathrm{\,g/L}$ ($1.7 \mathrm{\, M}$) ), a lower resistance is observed, with a deviation from the power law $S_{membrane}^{-1/2}$. Furthermore, as shown on Figure \ref{resistance}A for solution 1, increasing the length of the reservoirs by a factor 4 barely affects the resistance, showing that the measurement is not sensitive to the reservoir.

In a second step, the cell resistance is measured without membrane, but with the masking windows (in this case, we keep the notation $S_{membrane}$ to describe the opening area in the masking windows). Results are shown in Figure \ref{resistance}B for  solutions 1, 2 and 3. The three set of measurements show a decrease of the resistance with the opening area with a power law $S_{membrane}^{-1/2}$ over 4 decades of area. More precisely, for the solutions 1 and 2, the measured resistance is the same with and without membrane. The measurements with solution 3 show the same scaling law as for solution 1 and 2, on the contrary to the measurements with a membrane. %For every concentration, the resistance versus the area of the masking-window opening area $S_{membrane}$ yields a power law $S^{-1/2}$ 

%In a second step, the same measurements are performed with a Nafion membrane for the same concentrations (still symmetrical, i.e. without concentration ratio) and the same areas of masking-window opening. As shown in Figure \ref{resistance}, for concentrations $c_1$ and $c_2$ these measurements yield exactly the same results as without the membrane, and hence the same scaling: a power law $S_{membrane}^{-1/2}$. This scaling is a very good agreement with the measurements of Lni et al. \cite{Lin2023Essence}. For large concentration ($c_3$), a higher resistance is observed in the presence of the membrane. Furthermore, increasing the length of the reservoirs by a factor 4 barely affects the resistance, showing that the measurement is not sensitive to the reservoir.

%\color{JungleGreen} Peut etre mentionner le choix d'une résistance de membrane indépendante de la concentration qui explique la différence avec/sans membrane à haute concentration et petite surface (à enlever de la légende de la figure 3 pour alléger ?)
%\color{black}

\subsubsection{Theoretical modeling}

%\color{orange}
%Je pense mais je me trompe peut etre que cela serai plus clair si on exprimait la résistance avace tous les termes et pas seulement le terme en raciend e S
%\color{black} 

These results seems in contraction with the textbook usual equation of resistance $R$ which states:
\begin{equation}
    R_{membrane}=\dfrac{1}{\sigma_{m}} \cdot \dfrac{e_m}{S_{membrane}} \label{R_bulk}
\end{equation}
(with $e_m$ the membrane thickness, $\sigma_m$ the conductivity of the membrane, and $S_{membrane}$ its surface) which would yield a power law $R_{membrane}\propto S_{membrane}^{-1}$.

\begin{figure}[h]
\centering
    \includegraphics[width=1\linewidth]{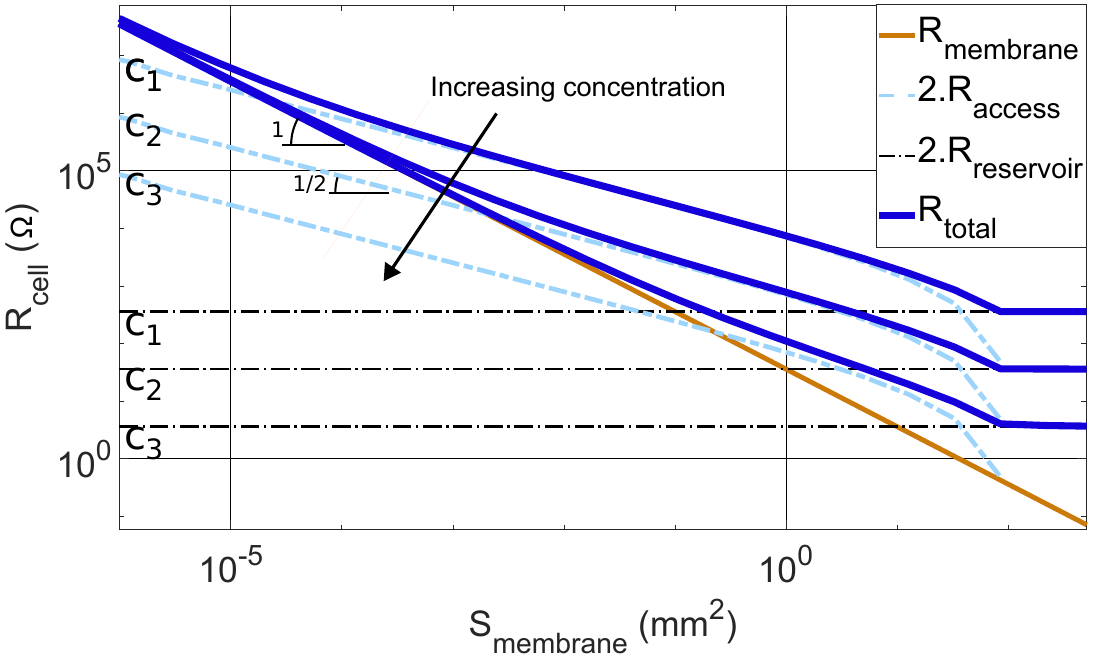}
    \caption{Evolution of the resistance with respect to the membrane size according to the developed model (with a perfectly selective membrane). The system is theoretically described by 5 resistances in series with different scaling with size. Curves are calculated for a symmetrical salt concentration $c_1$, $c_2$ and $c_3$. The membrane resistance (${R_{membrane}}$) is considered independent of the salt concentration, based on Avci \textit{et al}. \cite{Avci2020Energyharvesting}.
    \label{resistance_th}}
\end{figure}

Green \textit{et al}. \cite{Green2014EffectGeometry,Green2016Interplay} have shown that for a perfectly selective nanopore and for a non-selective nanopore connected to two reservoirs, the total resistance can be split in 5 as presented in Figure \ref{resistance_th}: the resistances of the two reservoirs on both sides, the resistance of the nanopore, and the access resistances at the two reservoir/nanopore connections:

\begin{equation}
    R_{cell}= R_{res, left}+R_{access, left} + R_{nanopore}+R_{access, right}+R_{res, right} \label{Rtotal}
\end{equation}

The reservoir resistance is $ R_{res, k} = \dfrac{1}{\sigma_i}\cdot \dfrac{L_{res,k}}{S_{res,k}}$ ($k=left,\, right$), with $\sigma_i$ the electrolyte conductivity ($i\in {1,2,3}$), $L_{res,k}$ the lengths of the reservoirs, and $S_{res,k}$ their cross-section areas. The nanopore resistance writes $ R_{nanopore} = \dfrac{1}{\sigma_{nanopore}}\cdot \dfrac{L_{nano}}{S_{nano}}$, with $\sigma_{nanopore}$ the conductivity of the nanopore (which takes into account both the co-ions and counter-ions), $L_{nano}$ its length, and $S_{nano}$ its cross-section area. While the nanopore from the initial model and the present membrane are different in geometry, they share the same permselective trait and therefore we treat them as they same and denote both as "\textit{membrane}", and use Eq \eqref{R_bulk}.

The access resistances originate from the mismatch between the two section areas $S_{res, k}$ and $S_{membrane}$ ($S_{nano}$ in the original paper), which imposes a convergence/divergence of field lines. There is no simple equation in the general case, but it has been modeled for arbitrary geometries by Green and collaborators \cite{Green2014EffectGeometry,Sebastian2023ElectricalCircuit}, and a simple equation has been proposed for an infinite reservoir ($S_{res,k}/S_{membrane}\rightarrow \infty$) and a circular nanopore  by Hall \cite{Hall1975AcessResistance}:

\begin{equation}
    R_{access, assymptotic}= \dfrac{1}{\sigma_i} \cdot \dfrac{\sqrt{\pi}}{4 \sqrt{S_{membrane}}} \label{R_nano}
\end{equation}
%In this model, while the reservoir and nanopore resistances scales in $S_{reservoir}^{-1}$ and $S_{nanopore}^{-1}$ respectively, the access resistances scale in $S_{nanopore}^{-1/2}$. These access resistances have been well known for decades in the case of nanopores: it originates from the mismatch between the two section areas $S_{res, left/right}$ and $S_{nanopore}$, which imposes a convergence/divergence of field lines. This has been modeled for an infinite reservoir ($\frac{S_{reservoir}}{S_{nanopore}}\rightarrow \infty$) and a circular nanopore by Hall \cite{Hall1975AcessResistance}:

As explained by Green \textit{et al.}\cite{Green2014EffectGeometry,Green2016Interplay}, the origin of the access resistances is purely geometrical, and it exists anytime $S_{membrane} \neq S_{reservoir}$.

The model of Green can thus be applied to the current system by replacing the nanopore by the membrane. The selectivity of the membrane is assumed to be high enough to allow for the decomposition of the total resistance in five as explained above. The access resistance therefore depends on the membrane area and shows an asymptotic scaling $R_{access}\sim \dfrac{1}{\sigma \sqrt{S_{membrane}}}$ for $S_{membrane}/S_{res} \ll 1$. The nanopore resistance becomes the membrane resistance described by Eq \eqref{R_bulk}. Finally, the reservoir section considered to model our system is the section of the electrodes.

%\begin{figure}[h]
%\centering
%    \includegraphics[width=1\linewidth]{Green model C = 10 coloré.pdf}
%    \caption{Alternative fig 3
%    \label{resistance_th_coloré}}
%\end{figure}

%Equation \eqref{R_bulk} can be used to calculate the nanopore resistance (which represents the membrane resistance in our case, and scales in $S_{nanopore}^{-1}$) and the reservoir resistance (independent of $S_{nanopore}$), and Equation \eqref{R_nano} for the access resistances. 

Hence, once the Nafion membrane is assumed to be perfectly selective, the 5 resistances of Eq \eqref{Rtotal} are known with the Nafion resistivity as only adjustable parameter, and sum up to yield the total resistance. Nafion-resistivity measurement from literature show poor variations with concentration \cite{Avci2020Energyharvesting}. The resistance is therefore chosen independent of concentration for simplicity. The best agreement between the model and experiments is found for a Nafion resistivity $R_{membrane}\cdot S_{membrane}=0.35 \mathrm{\, \Omega \cdot cm^2}$. It is reasonnable agreement with the values found in literature, considering the variability between the different reported values\cite{Lindheimer1987AstudyofNafion,Lehmani1997IonTransportInNafion,Avci2020Energyharvesting}.

%$\dfrac{e_m}{\sigma_m}=0.36 \mathrm{\, \Omega \cdot cm^2}$

As shown in Figure \ref{resistance_th}, the total resistance versus membrane area shows a power law $S_{membrane}^{-1/2}$ which indicates that the total resistance is dominated by the access resistances in the range of membrane surface considered here. This explains why the total resistance does not depend on the presence of the membrane, except at high concentration and small membrane area when the membrane resistance begins to be of the same order as the access resistance. This also explains why the total resistance does not depend on the reservoir length.

This model describes very well the experimental results for the different areas and solutions, as visible in Figure \ref{resistance}. This indicates that in this simple usual configuration, the resistance is dominated by the access resistance. As a consequence, the total conductance ($1/R_{total}$) does not scale linearly with the membrane surface.

The modeling on a broader range of membrane areas (Figure \ref{resistance_th}) shows that at very small membrane areas, the resistance should be dominated by the membrane resistance, and that on the opposite, when the membrane area becomes as large as the reservoir area, the total resistance should be dominated by the reservoir area.
% slope -1/2

% cite green and compare

%% continue from here

\subsection{Power}
\label{powerSec}

Now that the effect of membrane size is understood for homogeneous cells, asymetric configurations can be investigated. To this aim, different concentrations are introduced in the left and right reservoir. One always contains solution 1, and the other contains solution 2 or solution 3. This introduce a concentration ratio of 10 or 100 between the two sides of the membrane. Due to the selectivity of the membrane, this creates a potential difference, and a current if the electrical circuit is closed by a resistor.

The power that is injected in an external load resistance $R_{load}$ depends on the internal resistance $R_{cell}$ and the open-circuit voltage $E_{OCV}$  through the voltage divider: 
\begin{equation}
    P(R_{load})=E_{OCV}^2\cdot \dfrac{R_{load}}{(R_{load}+R_{cell})^2} \label{P_divider}
\end{equation}

The maximum power is then reached when $R_{load}=R_{cell}$:
\begin{equation}
    P_{max}=P(R_{load}=R_{cell})=\dfrac{E_{OCV}^2}{4R_{cell}} \label{Pmax}
\end{equation}

\begin{figure*}[t]
\centering
    \includegraphics[width=1\linewidth]{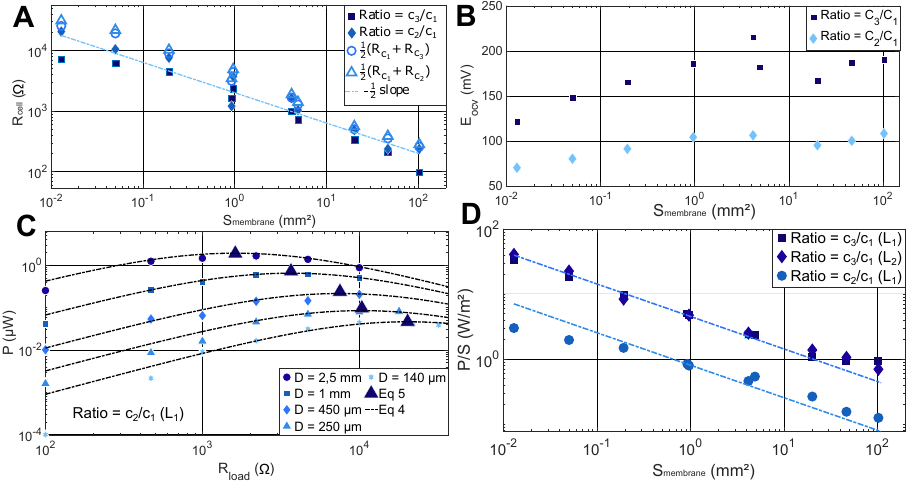}
    \caption{Measured power when the membrane is placed between reservoirs of different salinity. (a) Measured cell resistance ($R_{cell}$) for a varying $S_{membrane}$ for salt ratio of 10 and 100 (solid shapes). Open shapes are calculated from Fig \ref{resistance} assuming that the ratio configuration is half of $c_1$ resistance and half of $c_3$ resistance. Light blue dashed line is a $-1/2$ slope guide for the eyes. (b) Variation of the open-circuit voltage ($E_{ocv}$) with the membrane area ($S_{membrane}$): the OCV does not vary for large membranes, but drops at small areas. (c) Generated output power ($P$) as a function of the load resistance ($R_{load}$) for a salt ratio of 10: the maximum defines the maximum power density, in good agreement with the values calculated from Eq \eqref{Pmax} (solid triangles). Dashed line is the curve predicted by the cell resistance and OCV measurement from Eq \eqref{P_divider}. (d) Maximum power density ($P_{max}/S_{membrane}$) vs $S_{membrane}$ shows a power-law decrease with exponent $-1/2$. Measurements have been done for a salinity ratio of 10 and 100. The later have been measured for two electrode-membrane distances ${L_1}$ and ${L_2}$, with the same result.
    \label{power}}
\end{figure*}

The maximum power measurement can therefore be found by measuring the power for various $R_{load}$ (small symbols in Figure \ref{power}C), or by measuring $R_{cell}$ (Figure \ref{power}A) and $E_{OCV}$ (Figure \ref{power}B), yielding the power shown by the dashed lines and large triangles in Figure \ref{power}C).

%\color{brightmaroon}

Measurement of the open-circuit voltage (OCV) $E_{OCV}$ (Figure \ref{power}B) shows a significant variation with the membrane area: decreasing the membrane size induces a decrease of the OCV. This results is very surprising since for given solutions, the OCV is expected to be a function of the membrane selectivity only, and hence independent of the resistance and membrane size. Although the author have no clear understanding of this effect so far, these results may be related to the effect observed by Yazda \textit{et al.}\cite{Yazda2021HighOsmotic} for solid-state selective nanopores: they observed than when nine 4-nm wide nanopores come closer two each other, the OCV drops below a 500 nm distance. 

This effect could be due to the concentration field close to the membrane which arise from the very low (but non zero) osmotic flux of ions through the membrane (due to the non-ideal selectivity of the membrane), and which is called concentration polarization. This hypothesis is coherent with the observation that thicker masking windows leads to stronger decrease of the OCV, as shown in SI. %This would explain why the resistance is lower than expected in gradient configuration in the same range of membrane surface, as visible in Figure \ref{power}A (for $S_{membrane}< 10^{-1} \mathrm{\, mm^2}$): if a small quantity of ion from the concentrated electrolyte can cross the membrane, it will increase the concentration in the vicinity of the membrane in the diluted side. 

Measurements of the cell resistance are displayed on Figure \ref{power}A. 
Cell resistance measurements are shown in figure \ref{power}A. 
The values measured when the two tanks contain two different salinities (solution 1 and 2 or solution 1 and 3) are close to half the sum of the resistances of the two tanks containing the same solution. This means that cell resistance varies as $S_{membrane}^{-1/2}$ at least for membranes with surfaces greater than 10$^{-1}$ mm$^{2}$. This means that the ion-concentration profile is not very different in the two cases, or at least that the concentration profile in the dilute zone is not greatly affected by contact with a more concentrated tank. Remember that the cell resistance is largely controlled by the resistance of the compartment containing the more dilute solution. 
We note that this agreement is less good when the membrane surface is smaller. This situation corresponds to the case where the open circuit potential drops and reaches values less than 70 mV in the situation of solution 1 and 2  or 150 mV in the situation of solution 1 and 3. 
In those situations, it is reasonable to assume that the previous remark on the concentration profile in the diluted zone is not valid, and that it is affected by the fact that the diluted reservoir is in contact with a concentrated reservoir. 
Note that the resistance is 4 times smaller than the half-sum of the resistances, which implies a significant change in the concentration profile.
This indicates even more  that the measurement of the membrane selectivity could also be affected by size effects and the cell geometry, which would also hinder the comparison of data from literature. It is hence difficult to deduce the selectivity of the Nafion membrane from these experiments.

The power density (power per unit area of membrane $P_{max}/S_{membrane}$) is then plotted as a function of the membrane area $S_{membrane}$ in Figure \ref{power}D and it appears that the power density does depend on membrane area. The relation between power density and membrane area yields a power law with an exponent $-\frac{1}{2}$. From Equation \eqref{Pmax}, it is clear that the result from section \ref{resistanceSec} ($R_{cell}\propto S_{membrane}^{-1/2}$, which is still valid in the asymetric configuration when surface membranes are greater than 10${-1}$ mm$^{2}$ as visible in Figure \ref{power}A), yields a scaling law for the maximum power: $P_{max} \propto S_{membrane}^{1/2}$. From this, it is straightforward that the power density has a scaling law $P_{max}/S_{membrane} \propto S_{membrane}^{-1/2}$. For membrane size values smaller than 10$^{-1}$ mm$^{2}$, it seems that the scaling in $S_{membrane}^{-1/2}$ is still valid. This is due to a compensation phenomenon that we can't yet explain by the models. In this zone, the resistance is lower, but the circuit potential is not constant, but falls relative to the reference value.

As visible in Figure \ref{power}D, two sets of measurements performed on reservoirs of different lengths ($L_1$ and $L_2$) yield exactly the same results. This proves that the deviation from the usual assumption of constant power density is indeed due to the access resistance which dominates the total resistance of the device. This modeling also explains the trend observed by previous work from literature \cite{Pan2023TowardScalable,Lin2023Essence}. 

These results show that the size of the membrane does have an impact on the maximum power density that can be extracted: without precise analysis of the equivalent electrical circuit or a universal measurement protocol, the maximum power density should not be considered as an intrinsic parameter of the membrane, and is therefore not scalable from small scale experiments to larger scale.

\section{Discussion}
\label{DiscussionSec}
These results raise several difficulties for membrane characterisation, literature analysis, and scale up. 
% What is a good measurement geometry : fully 1D  : thickness always thinner that membrane diameter and electrode diameter.

First, this implies that membrane characterization must be done in carefully-chosen geometry to ensure that the dominating resistance (among the reservoir, access, and membrane resistances) is the one of interest. To probe the membrane, the access resistance and the reservoir resistance must be lower lower than the membrane resistance. This can be achieved by using electrode of the same size as the membrane, and as close as possible, possibly in contact with the membrane. 

%\color{JungleGreen} peut-être qu'on peut ajouter un graphique qui montre les courbes de Green dans un cas où on arrive à mesurer la résistance de membrane ?\color{black}

%Second, for literature analysis, it is necessary to know what resistance dominates the setup, to be sure that measurements can be compared. If the resistance is dominated by the access resistance, power measurements mostly measure the electrolyte conductivity and membrane selectivity. However, to have a critical analysis on literature results, it is crucial to know the size of the membrane (usually indicated), but also the size of the electrodes and the distance between the electrodes and the membrane (usually not indicated). It is therefore of prime importance to communicate all these information in research reports.

Second, in this work, we have demonstrated that each of the three terms contributes to the overall response of the system. While conducting a literature survery, to our dismay, we discovered that it is often the case that experimental papers do not provide enough details on the geometric setup. Namely, what is the size of the membrane (usually indicated), but also the size of the electrodes and the distance between the electrodes and the membrane (usually not indicated). Thus, we emphasize to future experimentalists the importance of providing a detailed description, even in the supplementary material, of the setup, and all geometrical parameters.

%Finally, for the membrane resistance measurements to be useful for scale up, measurements must be done in a geometry similar to the geometry of interest, which is usually a stack of membranes and 100-$\mu$m spacers. Consequently, a similar geometry must be adopted for membrane characterization to get meaningful data in the optics of a later scale up.
Finally, we emphasize that our work has focused on a simplified system which has only one membrane, used to characterize membranes. In applied systems, there is a stack of membranes separated by spacer of typically $100 \mathrm{\, \mu m}$. We provide two strategies on how we believe the resistance and power densities measured on single membranes can be scaled up to predict their performance in applied systems (stacks). A first option is to keep membrane and electrodes separated by $L_{spacer}=100 \; \mu$m, and have roughly the same diameters for electrodes and membrane ($D_{electrode} \simeq D_{membrane}$), and have them significantly larger than 100 $\mu$m. A second option is to have smaller membrane, as far as $D_{electrode} = D_{membrane}$ and $L_{spacer} \ll D_{membrane}$. Doing so, the power density measured should be meaningful and scalable.

Due to this artefact, power densities measured in other configuration can yield very high values of power density \cite{Fu2019atomicallyThin,Wang2020HoleyGraphene,Yang2022Advancing,Liu2020PowerGeneration}, which are not relevant in the scope of large-scale energy harvesting.

%\color{orange}
%Sans etre trop virulent rappeler que certaines puissances raportées sont pour le moins hasardeuses.
%On garde un truc plus agressif pour l'éditeur.
%\color{black}

\section{Conclusion}

To conclude, this systematic study on the effect of the membrane size for reverse electro-dialysis shows that when the electrodes are larger than the membrane, the resistance does not scale inversely with the membrane area, but with its square root. This is well understood by taking into account access effects at the entrance of the membrane. It results that the power density depends on the membrane size. As a consequence, without being very careful on the electro-chemical cell design, measurements of power densities on small membranes are overly optimistic and cannot be scaled up to predict the power density in a membrane stack.

\small

\section{Methods}

\subsection{Salt solutions}
%Electrical resistance, voltage and power were measured using salt solutions or varying concentrations with specific activity ratio (and hence conductivity). This ensures a defined ratio of 10 or 100. To that end, a solution at $c_1 = 1 \mathrm{\,g/L}$ is made. The following solutions are prepared to reach 10 or 100 times the conductivity $\sigma(c_1)$.

%$\sigma(c_2) = 10 \cdot \sigma(c_1) \rightarrow c_2\simeq 10 \mathrm{\,g/L}$, 

%$\sigma(c_3) = 100 \cdot \sigma(c_1) \rightarrow c_3\simeq 100 \mathrm{\,g/L}$.

Salt solutions were prepared using high-purity potassium chloride (Sigma-Aldrich) and Milli-Q deionized water. $K^+$ and $Cl^-$ have almost the same diffusivity. %This symmetry simplifies the result interpretation and is therefore used instead of the classical NaCl ($Na^+$ and $Cl^-$ have different diffusivity \cite{Ryzhkov2018TheoreticalElectrolyteDiffusion}).

The conductivity were measured using a conductimeter (MU 6100H VWR) at room temperature for all the solution to prevent the activity to be modified by the endothermic mixing reaction.

\subsection{Membrane preparation}
%Nafion, brand, etc... conducvtivity Nafion \cite{Avci2020Energyharvesting}

% preparation

Nafion 115 membrane needs to be prepared to ensure its maximum performances and stability. Needed pieces were cut from the Nafion sheet (Fuel Cell Store) and placed in deionized water for at least 24 hours before use. This ensures that all the hydrophilic negatively charges ${SO^{3-}}$ sites of the polymer are filled with water. The hydrated salt is therefore able to cross it \cite{gouldSimulationStudyIon}.

%Different soaking solutions have been tried: deionized water, $1 \mathrm{\,g/L}, 10 \mathrm{\,g/L}$ and $100 \mathrm{\,g/L}$. 
Results are not affected when the soaking solution is of the same concentration than the solution used during the experimentation (in symmetrical condition without concentration gradient). %If the soaking solution is of higher concentration, the salt stuck inside the membrane is released during the measurement which increases the local conductivity. This produces inaccurate results and was therefore avoided. 
For concentration-gradient measurement, the membrane is prepared in salt solution of the low concentration ($1 \mathrm{\,g/L}$). %This ensures that the low concentration reservoir isn't polluted by the membrane salt. No measurable effect were observed for the high concentration reservoir.

\subsection{Electro-chemical cell}
% electrodes

An electro-chemical cell (Figure \ref{experimental_setup}) was designed and 3D printed (clear resin V4, Formlabs ${3B^+}$). The Nafion membrane with the two masking windows was clamped in a mounting bracket. It was then pressed between two O-ring in the middle of the cell by the two screwable reservoirs. From each side come the water inlet and outlet, the silver chloride electrode and the platinum electrode. Tubings and wires are connected to the cell using screwed mounting tip (Nanoport). Teflon tape is added when needed for the water sealing. Relevant sizes of the cell are listed in Table \ref{table_dimension}.

Silver electrodes are used as working electrodes and counter electrodes. When needed to solve electrode-polarization issues, platinum electrodes are used to measure the potential of the solution, while the current flows through the silver-chloride electrodes. %to measure the open-circuit potential $E_{OCV}$, which means the potential when the current is zero. However, when the current is non zero, the polarization at the electrode-electrolyte interface for concentrated electrolytes induces a potential drop at the electrode surface \cite{Bazant2005currentvoltage}. Accordingly, the silver-chloride electrodes cannot be used simultaneously to carry the current and to measure the potential. To this aim, platinum electrode are used to measure the potential of the solution, while the current flows through the silver-chloride electrodes.

\begin{table}
\renewcommand{\arraystretch}{1.2}
    \centering
    \begin{tabular}{|l|m{4 cm}|m{2.6 cm}|}
     \hline
     Name & Description & Size \\ [0.5ex] 
     \hline\hline
     ${D_r}$ & Inner reservoir diameter & 29 mm\\ 
     \hline
     ${L_r}$ & Length between the electrode and the membrane & ${L_1 = 3 mm}$ (default value, unless otherwise specified) ${L_2 = 13 mm}$\\
     \hline
     ${V_r}$ & Volume of the reservoir & 10 mL \\
     \hline
     ${D_e}$ & Diameter of the electrode & ${D_1 = 1 cm}$(default value, unless otherwise specified) ${D_2 = 2 cm}$\\
     \hline
     ${S_e}$ & Surface of the electrode & 1 cm² \\
     \hline
     ${D_w}$ & Diameter of the masking window & [140 µm; 10 mm]\\
     \hline
     ${S_{membrane}}$ & Surface of the masking window, equal to the surface of the useful membrane & [0,013 ; 103] mm²\\
     \hline
     ${e_m}$ & Membrane thickness & 100 µm \\
     \hline
    \end{tabular}
    \caption{Relevant sizes and geometrical parameters of the electro-chemical cell.}
    \label{table_dimension}
\end{table}

Silver chloride electrodes are made with a spiral-shaped 1mm-diameter silver wire (Figure \ref{experimental_setup} B). For each electrode, the outer diameter of the spiral is 1,5 cm and the total surface area is 1 cm². The silver wire was polished with sandpaper and cleaned with ethanol and deionized water. It was then placed in a beaker with highly concentrated NaCl solution. A carbon counter electrode allows to perform a chronopotentiometry measurement: a current of 15 mA is imposed while measuring the potential difference until it reaches 1,8 V. At this point, the electrode is completely covered with silver chloride. With a razor blade, part of this deposit is scratch to recover the silver. This ensures that both silver and silver chloride are in contact with the solution to allow the electrode to work as a faradaic electrode.

\subsection{Masking windows}
% discuss two window vs 1 window 

Masking windows were laser cut in a 125 µm thick Mylar sheet (Figure \ref{experimental_setup} A). % Because of the Mylar flexibility the total area needs to be close to the hole area to prevent from deformation. Therefore, 
Two sets of windows were prepared. One with holes between 140 µm and 2,5mm (Mylar external size is 3x3 mm) and the other between 1mm and 10mm (Mylar external size is 11x11 mm). 1mm and 2,5mm sizes are in both sets to ensure a reliable continuity of the data sets.

The free area of the window defines the effective area of the membrane. %It has been measured using pictures or microscope images. With ImageJ, a scaling, binarisation and boundaries detection is done to be able to calculate the real window area. 
The table in annex gives the average area of the two windows of the same intended size. All the windows are circular except for the 10x10 mm which is a square.

A quick study has been made to compare the results with a window only on one side of the membrane or on both sides. Because of the isotropic properties of Nafion\cite{Ma2006isotropicNafion}, the ion conduction isn't in straight line inside of the membrane. Therefore, when only one window is used, the flux in the membrane will take place in a volume which have a cross section larger than the hole in the masking window. To be closer to the ideal case of a membrane of the exact size, two windows are placed on each side of the Nafion sheet. The conduction is still not completely straight. But the deviation might be lower than the membrane thickness and hence smaller than the windows diameters.

%The two windows were aligned into a square hole in the membrane mounting bracket. 
Due to the laser cutting precision and the handling variations, some misalignment between the windows could happen. It was statistically measured to be under 100 µm. Thanks to the isotropic properties of pristine Nafion \cite{Ma2006isotropicNafion}, this positioning error could not induce an increase of membrane effective thickness larger than 41$\%$ in the worst case. This would increase the membrane resistance value by the same ratio. We assume that it has a very minor effect in our study.%, therefore it hasn't been optimized further.

\subsection{Electrical circuit}

Electrical measurements were performed using a potentiometer (BioLogic SP 300). This device has 5 wires per channel: Sensing electrode 1 (S1), Sensing electrode 2 (S3), reference electrode (S2), Working electrode (P1) and counter electrode (P2). The reference electrode S2 is always connected to the sensing electrode S1.

When silver-chloride electrodes are crossed by a current, a complex concentration polarization and capacitive double layer can take place around them \cite{Bazant2005currentvoltage}, which increase the measured resistance. This effect can be avoided by using 4 electrodes (Platinum electrodes in addition to the silver-chloride electrodes), or by using configurations when the electrode polarization is either absent or negligible. In the first configuration (4-electrode setup), the two platinum wires are used as sensing electrodes (connected to S1, and S3) (see SI for more detailed information). No current is going through them, they only measure the potential drop between the two reservoirs. The two silver-chloride electrodes (connected to P1 and P2) are used to inject current through the device. Therefore, the sensing and the current conduction are separated and there isn't any polarization issue on the sensing electrodes. This measurement setup is used to measure the cell resistance with or without salt ratio.

The second configuration (2-electrode configuration) is used to measure the open circuit voltage (${E_{OCV}}$) of the cell with a concentration gradient and its response to an external load resistor (${R_{load}}$). In those cases the voltage measurements are performed on the working- and counter electrodes. During ${E_{OCV}}$ measurement, no current is going through the circuit, and hence no capacitive double layer is created on the electrode surface. During the measurements with varying ${R_{load}}$, the high resistance of the dilute reservoir (which is highly resistive and is too dilute to create a significant capacitive double layer) dominates the overall resistance: consequently in this specific case, the variation of the high-concentration reservoir due to the polarization do not affect the total resistance. Therefore, in both cases the previously-mentioned issue with silver chloride electrodes has no effect on the measurement.

\subsection{Measurement protocol}

For all the presented measurements, salt solutions, membrane, electro-chemical cell, masking windows and electrical circuit are prepared as explained before. The resistance measurement is made in the 4-electrode configuration (see Supp Mat for more details). Each side of the cell is filled with water using a peristaltic pump (LongerPump BT 100-1L), either with the same salt concentration on both sides (symmetric condition), or with a different concentration (gradient condition). In this last configuration, the ratio value is defined as the activity ratio (equal to conductivity ratio) between both reservoirs (Ratio = 10 or 100). When there is no air in the cell any more, the pump is stopped. To ensure that there isn't any pressure differences and parasite flows inside of the cell, measurements start only after a relaxation time of a few minutes. Once a steady state in potential is reached, the measurement begins.

To measure the resistance, the potentiostat applies a constant current ${I_{mes}}$ during 10s to the cell through the silver chloride electrodes (P1 and P2). The response is a potential increase (${E_{mes}}$) measured by the platinum wires (S1 and S3). This operation is done with 3 different ${I_{mes}}$ values (100 µA, 50 µA and 10 µA) with a relaxation time of 30 s between each measurmeents. The linear regression of the curve ${E_{mes}}$-${I_{mes}}$ gives the cell resistance ${R_{cell}}$. ${R_{cell}}$ is measured with this protocol for all windows sizes, with or without concentration gradient.

For the study with a concentration gradient, the value of ${E_{OCV}}$ is needed to calculate the output power. The 2-electrode configuration is used (potential measured through the silver chloride electrodes). As for the resistance measurement, the device is filled with water and left to balance without water flux. The potential ${E_{OCV}}$ is then measured until stable for several minutes. 

This same configuration is used to measure the power dependency to the outer load resistor ${R_{load}}$. For each value of ${R_{load}}$, the circuit is closed and the potential drop $E_R$ across the load resistor is measured. From the potential drop across the resistor, the power $P$  generated by the circuit for this ${R_{load}}$ value is calculated : $P=E_R^2/R_{load}$.

% OCV and resistance decoupled

\section*{Acknowledgment}
The authors warmly thank Prof. Yoav Green for the fruitful discussions about the theoretical modeling of our experimental results.

TD acknowledges funding from the Institut Pierre-Gilles de Gennes (laboratoire d'excellence PSL).

\section*{Author Contributions}
TD, CT, AC designed the experiments. TD performed the measurements. TD, CT, AC analyzed and discussed the data. TD, CT, AC discussed the modeling. TD, CT, AC wrote, reviewed and edited  the manuscript.

\section*{Conflicts of interest}
There are no conflicts to declare.

%%%END OF MAIN TEXT%%%

%  For footnotes in the main text of the article please number the footnotes to avoid duplicate symbols. e.g.  \footnote[num]{your text} the corresponding author \ast counts as footnote 1, ESI as footnote 2, e.g. if there is no ESI, please start at [num]=[2], if ESI is cited in the title please start at [num]=[3] etc. Please also cite the ESI within the main body of the text using \dag.

% The \balance command can be used to balance the columns on the final page if desired. It should be placed anywhere within the first column of the last page.

% \balance

% If notes are included in your references you can change the title from 'References' to 'Notes and references' using the following command:
% \renewcommand\refname{Notes and references}

%%%REFERENCES%%%
\scriptsize{
\bibliography{refintegrated} %You need to replace "rsc" on this line with the name of your .bib file
\bibliographystyle{unsrt} } %the RSC's .bst file

\end{document}